\documentclass[structabstract]{aa}

\usepackage{txfonts}
\usepackage{natbib}
\usepackage[]{graphicx}

\begin{document}

\title{A Systematic Examination of Particle Motion in a Collapsing Magnetic Trap Model for Solar Flares}

\author{K. J. Grady \and T. Neukirch \and P. Giuliani}

\date{ Received / Accepted}

\institute{School of Mathematics and Statistics, University of St.~Andrews, St~Andrews KY16 9SS, UK}

\abstract
{It has been suggested that collapsing magnetic traps may contribute to accelerating particles to high energies during solar flares.}
{We present a detailed investigation of the energization processes of particles in collapsing 
magnetic traps, using a specific model. We also compare for the first time the energization processes in a symmetric and an asymmetric trap model. }
{Particle orbits are calculated using guiding centre theory. 
We systematically investigate the dependence of the energization process on initial position, initial energy and
initial pitch angle.}
{We find that in our symmetric trap model particles can gain up to about $50$ times their initial energy, but that 
for most initial conditions the energy gain is more moderate.
Particles with an initial position in the weak field region of the collapsing trap and with 
pitch angles around $90^\circ$ achieve the highest
energy gain, with betatron acceleration of the perpendicular energy the dominant energization mechanism.
For particles with smaller initial pitch angle, but still outside the loss cone, we find the possibility of a 
significant increase in parallel
energy. { This increase in parallel energy can be attributed to the curvature term in the parallel equation of motion and the associated energy gain
happens in the  center of the trap where the field line curvature has its maximum.}
We find qualitatively similar results for the asymmetric trap model, but with smaller energy gains and a larger number of particles escaping from the trap.}{}



\titlerunning{ Particle Motion in Collapsing Magnetic Trap Model}
\authorrunning{Grady, Neukirch \& Giuliani}

\maketitle

\section{Introduction}

One of the most important open questions in solar physics is to identify the mechanisms 
by which  a large 
number of charged particles are accelerated to high energies during solar flares.
While there is general consensus that the energy released in solar flares is previously 
stored in the magnetic field,  the physical processes by which this energy is converted 
into bulk flow energy, thermal energy, non-thermal energy and radiation energy are still a 
matter of debate \citep[e.g.][]{miller:etal97,aschwanden02,neukirch05b,neukirch:etal07,krucker:etal08,aschwanden09}.
Based on observations, in particular of non-thermal high-energy (hard X-ray and $\gamma$-ray) 
radiation, it has been estimated that a large fraction of the released magnetic 
energy (up to  the order of 50 \%)  is converted into non-thermal energy in the 
form of high energy particles \citep[e.g.][]{emslie:etal04,emslie:etal05}. 

Possible particle acceleration mechanisms that have been suggested include
direct acceleration in the parallel electric field associated with the reconnection 
process, stochastic acceleration by turbulence and/or wave-particle resonance, 
shock acceleration or acceleration in the inductive electric field of the 
reconfiguring 
magnetic field \citep[see e.g.][for a detailed discussion and further references]{miller:etal97,aschwanden02,neukirch05b,neukirch:etal07,krucker:etal08,zharkova:etal11}. 
As none of the 
proposed mechanisms can fully explain the high-energy particle fluxes within the framework of the 
standard solar flare thick target model, alternative acceleration scenarios have been proposed \citep[e.g.][]{fletcher:hudson08,birn:etal09, brown:etal09}.

It has been suggested \citep[e.g.][]{somov92,somov:kosugi97} that the rapid relaxation of magnetic field lines that have been newly reconnected, 
but outside the actual reconnection region, could contribute to the acceleration of particles. Charged particles could be trapped 
within the dynamically relaxing magnetic fields, which form a collapsing magnetic trap (CMT from now on).
 Based on the conservation 
 of adiabatic constants of motion, the kinetic energy of the particles trapped in a CMT could increase due to the betatron effect, 
 as the magnetic field strength in the CMT increases, and due to first-order Fermi acceleration, as the distance between the 
 mirror points of particle orbits decreases due to the shortening of the field lines.
Evidence of post-flare field line relaxation (field line shrinkage) has been found in observations by Yohkoh \citep[e.g.][]{forbes:acton96} and 
Hinode \citep[e.g.][]{reeves:etal08a}. Field line relaxation in solar flares can be compared to the  dipolarisation phase of magnetospheric 
substorms, which is believed to play a major role in particle acceleration during substorms
\citep[e.g.][]{birn:etal97,birn:etal98,birn:etal04}.  A more general comparison of flare and substorm/magnetotail phenomena 
based on observations has recently been presented by \citet{reeves:etal08b}.
Another, albeit slightly different, particle acceleration mechanism, which also relies on the relaxation of magnetic field lines is the
shrinkage of magnetic islands (plasmoids) that has been discussed, for example by \citet{drake:etal06} and \citet{karlicky:barta07}.

Using very basic CMT models, some of the fundamental properties of the particle acceleration process in CMTs have been investigated by Somov and co-workers \citep[e.g.][]{bogachev:somov01,bogachev:somov05,bogachev:somov09,kovalev:somov02,kovalev:somov03a,kovalev:somov03b,somov:bogachev03}. 
This includes, for example, the relative efficiencies of betatron and Fermi acceleration,  the effect of collisions, the role of velocity anisotropies  and the evolution of the energy distribution function in a CMT. \citet{li:fleishman09} used the results of \citet{bogachev:somov05} to determine the gyrosynchroton radio emission
that is to be expected from a collapsing magnetic trap. While these calculations are extremely useful for first estimates, they are based on a number of simplifying
assumptions about both the CMTs and the particle orbits. Specific models are necessary for gaining a more detailed understanding of the processes in CMTs,
with the disadvantage that some of the results will become model-dependent.

A  number of previous papers has followed this line of investigation.
Using a simple model for time evolution of the CMT magnetic field strength and a simplified equation of motion for the particles,
\citet{karlicky:kosugi04} investigated CMT properties such as particle acceleration, plasma heating and the resulting X-ray emission.
\citet{karlicky:barta06} used  an MHD (magnetohydrodynamic) simulation of a reconnecting current sheet to generate CMT-like electromagnetic fields to investigate 
particle acceleration using test particle calculations, in particular with a view to explain hard X-ray loop-top sources.  Some indication that CMTs might be relevant for X-ray loop top sources has been provided by \citet{veronig06}. \citet{aschwanden04} used a very simple time-dependent trap 
model to try and explain the pulsed time profile of energetic particle injection often 
observed during flares. Most recently, \citet{minoshima:etal10} presented the results of a calculation using numerical
solutions of the drift-kinetic equation for a CMT model based on a time-dependent 2D 
magnetic field suggested by \citet{lin:etal95} to interpret the motion of flare loops and ribbons in the framework of the standard
flare model.

\citet{giuliani:etal05} presented a general theoretical framework for more detailed analytical CMT models. This theoretical framework is based on general analytic solutions of the kinematic MHD equations, i.e. with given bulk flow profile.  The theory was developed by \citet{giuliani:etal05}  for 2D and 2.5D magnetic fields, but excluding flow in the invariant direction and has recently been extended to fully 3D magnetic fields and flows by \citet{grady:neukirch09b}.


\citet{giuliani:etal05} focus on the development of the theory and present just a few 
examples of model CMTs, together with a calculation of just a single example of a 
particle orbit based on non-relativistic guiding centre theory \citep[see e.g.][]{northrop63}. 
%
 In this paper we present a systematic investigation of test particle orbits for different initial conditions, firstly
by  using the same symmetric CMT model as in \citet{giuliani:etal05} and secondly by using a modified asymmetric CMT model. As in \citet{giuliani:etal05} we 
 will use the first order guiding centre theory. 
 We are in particular interested in the dependence of particle energy gain on 
 initial position in the trap, initial energy and initial pitch angle.
Another interesting question is whether the energy gain mechanisms predicted using 
adiabatic invariants can indeed be identified using the 
guiding centre orbits. 

The paper is structured in the following way. In section \ref{sec:basic} we summarise the 
basic theory and the models presented in \citet{giuliani:etal05}.
An overview of the dependence of particle orbits and energy on initial conditions for the symmetric CMT model is given 
in section \ref{sec:overview}. We take a more detailed look
at two particular particle orbits for the symmetric CMT model in section \ref{sec:mechanisms} to study the energy gain 
mechanisms in more detail.
In section \ref{sec:asymmetric} we the present results for an asymmetric CMT model.
We finish the paper
with summary and conclusions in section \ref{sec:summary}.

\section{Basic equations and CMT model}
\label{sec:basic}

 \citet{giuliani:etal05} develop their theory 
using the ideal kinematic MHD equations,
\begin{eqnarray}
\mathbf{E} + \mathbf{v} \times \mathbf{B} = \mathbf{0} \mbox{,} \label{ohm}\\
\frac{\partial \mathbf{B}}{\partial t} = - \nabla \times \mathbf{E} \mbox{,} \label{faraday}\\
\nabla \cdot \mathbf{B} = 0\mbox{.} \label{eq:solenoidal}
\end{eqnarray}
Under the assumption that all $z$-derivatives vanish, one can write the magnetic field
in the form
\begin{equation}
\mathbf{B}(x,y,t) = \nabla A(x,y,t) \times \mathbf{e}_z + B_z(x,y,t) \mathbf{e}_z,
\end{equation}
where $A(x,y,t)$ is the magnetic flux function. This form of the magnetic field automatically satisfies the solenoidal condition, Eq. (\ref{eq:solenoidal}).

Following \citet{giuliani:etal05} we will assume in this paper that both $B_z $ 
 and $v_z$ vanish \citep[for an extension of the theory to 3D, see][]{grady:neukirch09b}.
 Here the $x$-coordinate runs parallel to the solar surface, while the $y$-coordinate specifies
 the height above the solar surface.
 
The ideal Ohm's law (\ref{ohm}) then takes the form
\begin{equation}
\frac{\partial A}{\partial t} + \mathbf{v}\cdot\nabla A =0
\end{equation}
A CMT model is then defined by specifying the flux function $A$ at a specific time and a velocity field $\mathbf{v}(x,y,t)$. 
In the present paper we will use the flux function defining the magnetic field as $t\to\infty$. 
Instead of defining the velocity field directly, we specify a transformation from Eulerian to Lagrangian coordinates, 
which is equivalent to specifying the trajectories of plasma elements perpendicular to the magnetic field. 

In this paper we use the same CMT model as used by \citet{giuliani:etal05}. The final magnetic field is determined by the flux function
\begin{equation}
A_0=c_1 \arctan \left( \frac{y_0 + d_1}{x_0 + w}\right) - c_1 \arctan \left( \frac{y_0 + d_2}{x_0 - w}\right).
\label{AGiuliani}
\end{equation}
where $c_1$ is used to control the strength of the magnetic field and where $x_0$, $y_0$, $d_1$, $d_2$ and $w$ are considered to be normalised by length $L$, which
is the characteristic size of the trap.

The corresponding magnetic field is a loop between two 2D magnetic sources (line currents) separated by a distance $2w$ and placed below the photosphere
at $y_0=-d_1$ 
at $x_0=-w$ and $y_0=-d_2$ at $x_0=w$. 
The magnetic field generated by the flux function (\ref{AGiuliani}) is potential if regarded as a function of $x_0$ and $y_0$. This potential field
is the final field to which the CMT relaxes as $t \to \infty$. \citet{giuliani:etal05} used the values $d_1=d_2=1$ and $w=0.5$ in their paper, which
creates a magnetic loop which is symmetric with respect to $x=0$ at all times.

\begin{figure}
\resizebox{\hsize}{!}{\includegraphics{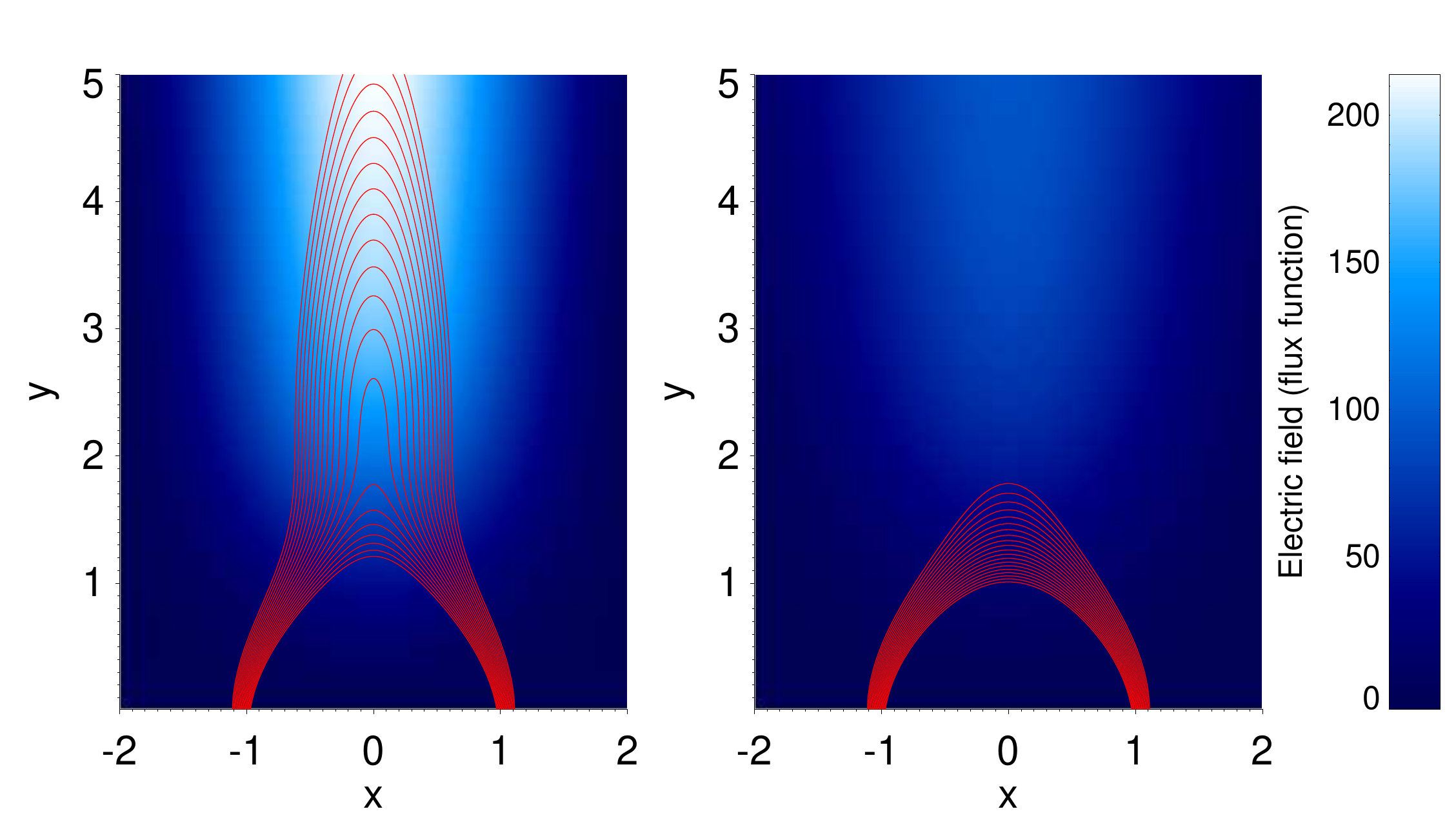}}
\caption{The magnetic field lines (red) and the electric field (blue colour scale) of the CMT model by \citet{giuliani:etal05} at the beginning (left panel) and at the end (right panel) of the collapse, corresponding to $95$~s in the normalisation used. Lengths are normalised to $\mbox{L} = 10^7$~m.}
\label{fig:cmtmodel}
\end{figure}

The coordinate transformation used by \citet{giuliani:etal05} is given by
\begin{eqnarray}
x_0&=&x \mbox{,} \\
y_0&=&(a t)^b \ln \left[ 1+ \frac{y}{(a t)^b}\right]
 \left\{ \frac{1+ \tanh[(y- L_v/L) a_1]}{2}\right\} \nonumber \\
& & +\left\{ \frac{1+ \tanh[(y- L_v/L) a_1]}{2}\right\} y \mbox{.}
 \label{ytransform2d}
\end{eqnarray} 
This transformation stretches the magnetic field in the $y$-direction above a height 
given by $L_v/L$, where the transition between unstretched and stretched field is controlled by the parameter $a_1$. In this paper we will use the 
same parameter values as \citet{giuliani:etal05}, namely $a=0.4$, $b=1.0$, $L_v/L=1$ 
and $a_1=0.9$. 
For simplicity, the transformation depends on time only through the function $y_0(y,t)$. This time-dependence lets the field collapse to the final field described above as for $t\to\infty$, $y_0$ tends to $y$.
An important feature of the transformation is that the foot points of magnetic field lines do not move during the collapse as for $y=0$ we have $y_0=0$ for all $t$.
In Fig.\ \ref{fig:cmtmodel} we show a plot of the magnetic field lines, and the electric field, at the start and at the end of the collapse.

For reasons of comparability, we also use the same normalisation as used by \citet{giuliani:etal05}, i.e.~the typical length scale of the trap is $\mbox{L} = 10^7$ m, the magnetic field is normalised by $0.01$ T ($100$ G) and the time scale for the collapse of the trap is $100$ s. We remark that these are rather conservative values, giving, for example, a typical field strength of $2\cdot 10^{-3}$ T ($20$ G) and below at the initial positions of the particles.
Stronger magnetic fields and smaller time-scales of CMT collapse are possible for solar flares and the consequences of varying the CMT parameters, and also the CMT model itself will be investigated in more detail in the future.

The particle orbits are calculated using first order non-relativistic guiding centre theory \citep[see e.g.][]{northrop63}
\begin{eqnarray*}
\left(\frac{m}{q}\right)\frac{d v_{\parallel}}{dt}=
E_{\parallel}- \frac{M}{q}\frac{\partial B}{\partial s}
+\left(\frac{m}{q}\right){\bf u}_{E}\cdot
\left( \frac{\partial {\bf b}}{\partial t} +
v_{\parallel}  \frac{\partial {\bf b}}{\partial s} +
{\bf u}_{E} \cdot\nabla{\bf b}
\right)
\end{eqnarray*}
\begin{eqnarray*}
\dot{\bf R}_{\perp} =& &\frac{\bf b}{B}  \times 
 \left\{
- {\bf E} + \frac{M}{q} \nabla B +
\frac{m}{q}
\left[
\right.\right.\\
& &
\left.\left.
v_{\parallel} \frac{\partial \bf b}{\partial t}
+
v_{\parallel}^2 \frac{\partial \bf b}{\partial s} +
v_{\parallel} {\bf u}_{E}\cdot\nabla{\bf b} +
\frac{\partial {\bf u}_{E}}{\partial t} +
v_{\parallel}\frac{\partial {\bf u}_{E}}{\partial s} +
{\bf u}_{E}\cdot \nabla {\bf u}_E
\right]
\right\}
\end{eqnarray*}
where $M=(1/2)m v_g^2 /B$  is the magnetic moment , which is regarded as constant in this approximation and thus fixed by the initial conditions, $v_g$ is the
gyro-velocity, ${\bf u}_{E}= ({\bf E}\times {\bf b})/B$,
${\bf b}={\bf B}/B$, ${\bf R}$ is the vector location of the
guiding centre, $v_{\parallel} = {\bf b}\cdot{\dot{\bf R}}$ and
$\dot{\bf R}_{\perp} = \dot{\bf R}-v_{\parallel}{\bf b}$ \citep[for evolution equations for the perpendicular and 
parallel energies, see][Eqs. (49) and (50)]{giuliani:etal05}.

In the present paper we only calculate electron orbits.
For all electron orbits presented in this paper the use of the guiding centre approximation is well justified, as, for example, the
typical ratio between the gyration timescale for electrons and the time-scale for the variation of the magnetic field of the CMT is of the order
$10^{-3}$ -- $10^{-4}$, and the differences between typical gyroradii and the MHD length scales of the CMT are also of this order.

\begin{figure}
\resizebox{\hsize}{!}{\includegraphics{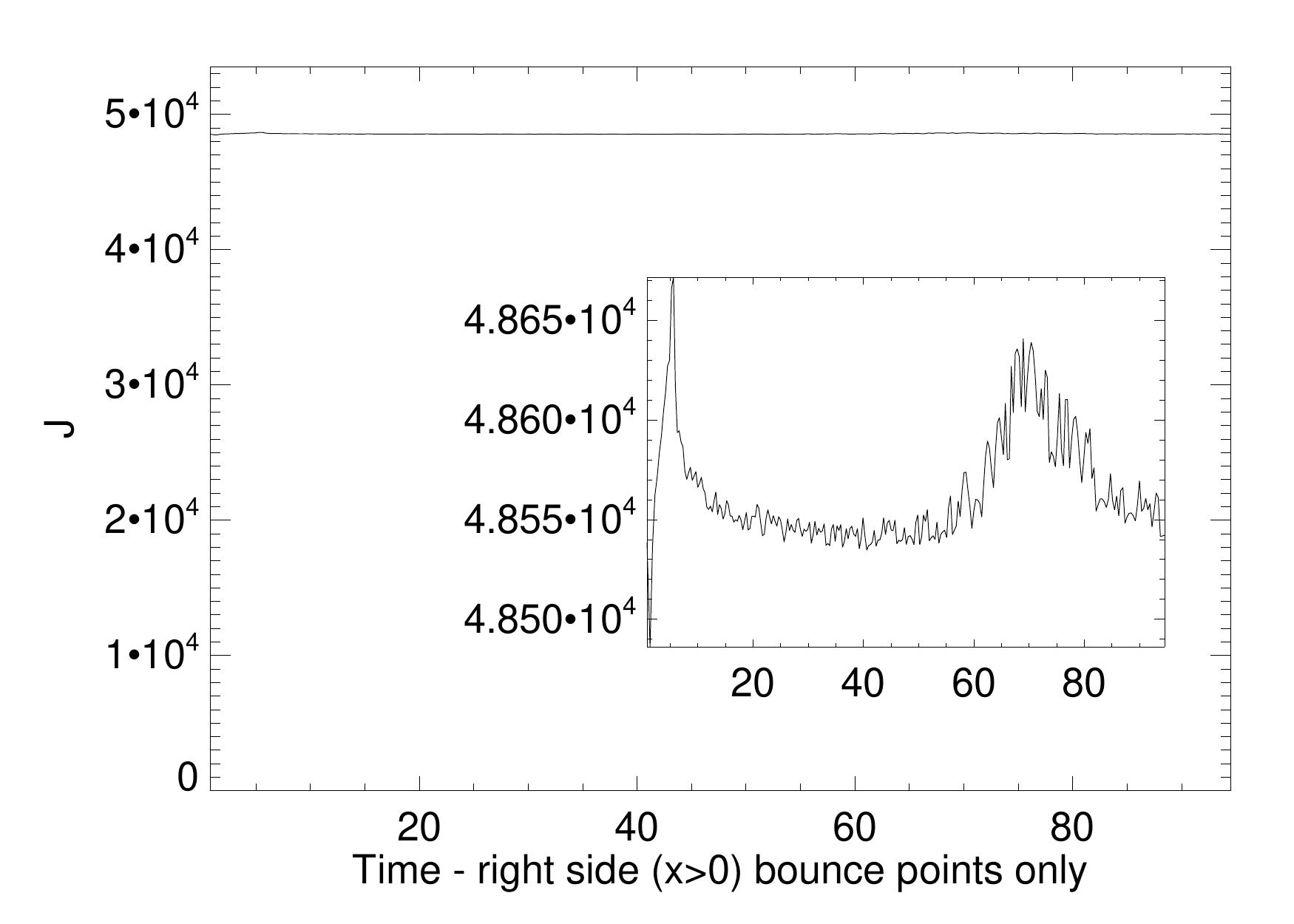}}
  \caption{Time variation of the longitudinal invariant $J$ for the electron orbit in \citet{giuliani:etal05}.}
\label{fig:j-invariant}
\end{figure}

The longitudinal adiabatic invariant
\begin{equation}
J=\oint m v_{\|} {\rm{d}}s \mbox{,}
\end{equation}
is a good invariant for the orbits studied in this paper. As an example we show in Fig.\ \ref{fig:j-invariant} the variation of $J$ for the orbit
discussed by \citet{giuliani:etal05}. We evaluated the path integral along the orbit between two consecutive
mirror points on the same side of the trap.
The variation of $J$ along the orbit is less than $0.5 \%$. This is typical for all orbits studied in this paper.

\section{Electron energy gains for varying initial conditions}
\label{sec:overview}

\subsection{Discussion of initial conditions}

In this section we investigate the influence of initial conditions on the energy gain of electrons in the CMT model of \citet{giuliani:etal05}. Generally, the initial position and initial velocity are varied. Regarding
the initial position, only the initial $x$ and $y$-values need to be varied, because the CMT model is invariant in the $z$-direction. 
Due to our use of guiding centre theory, we do not need to specify the complete 
initial velocity vector. In the present paper, we choose to specify the total initial kinetic energy and the initial pitch angle ($\alpha$) of the particles. Together with the initial position, this fixes the magnetic moment ($M$) 
of the 
particles. It also implicitly fixes the initial parallel ($E_\parallel = \frac{1}{2} m v_\parallel^2$) and perpendicular ($E_\perp = M B$) energies of the particle.

In the following we distinguish between particle orbits that have $y >0$ for all times (trapped particles) and particle orbits that eventually cross the lower boundary ($y=0$; escaping particles). For escaping
particles, the final energy and other quantities are recorded at the time of escape, i.e. when their orbit first reaches a value $y<0$, whereas for particles which remain trapped the corresponding values are recorded at the
final time of the calculation, i.e. when the trap is sufficiently relaxed.

\begin{figure}
\resizebox{\hsize}{!}{\includegraphics{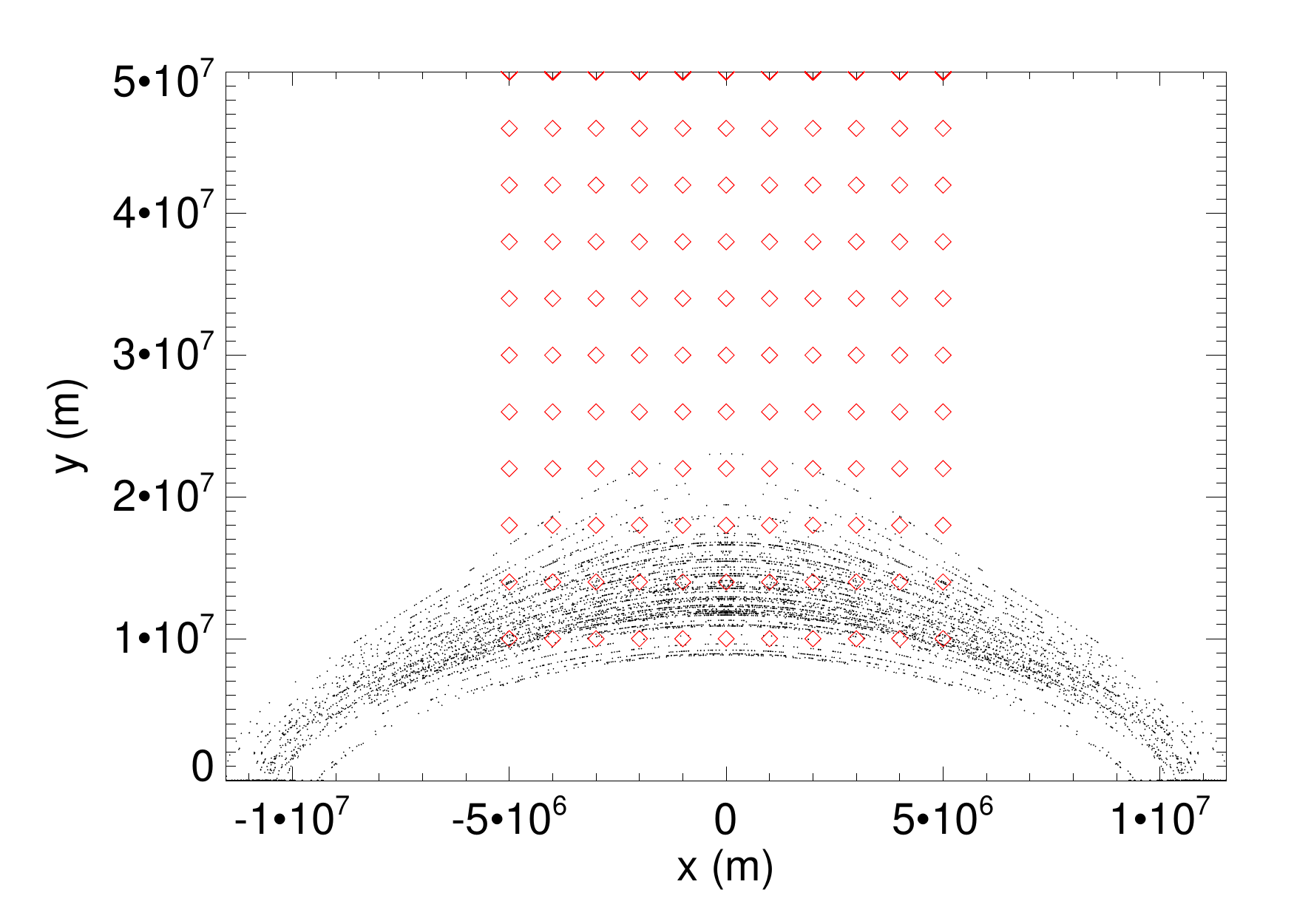}}
  \caption{Initial (diamond) and final (dot) positions for test particles.  We remark that we only represent individual orbits, which are not representative of, for example, particle densities. A stronger concentration of particles in the 
  centre of the trap at the initial time would lead to an even higher density of energetic particles trapped at the loop top at the final time.}
  \label{initialpos}
\end{figure}

To study the effects of varying the initial conditions, we use a grid of  $11$ by $11$ equidistantly spaced initial positions for $-0.5 \mbox{ L} \le x \le 0.5 \mbox{ L}$ and 
$1 \mbox{ L} \le y \le 5\mbox{ L}$ (see diamond shaped symbols in Fig.\ \ref{initialpos}). For each initial position we calculate particle orbits for 
$11$ equally spaced values between $5$ keV and $6$ keV for the initial energy and 
$10$ values for the initial pitch angle between $13^\circ$ and $163^\circ$.
In Fig.\ \ref{initialpos} we also show the final positions of the particles remaining in the trap as dots. 
Particles which start at positions further away from the centre of the trap ($x=0$) are more likely to escape quickly even for initial pitch angles relatively close to $90^\circ$, often without mirroring, whereas particles starting close to the centre are more likely to remain trapped. The reason is that outside the main region of the CMT the magnetic field strength does not vary as much as inside the CMT and mirroring is less likely to occur, i.e. the particles start with initial conditions
putting them inside the local loss cone.

\subsection{Dependence of energy gain on initial position}



\begin{figure}
\resizebox{\hsize}{!}{\includegraphics{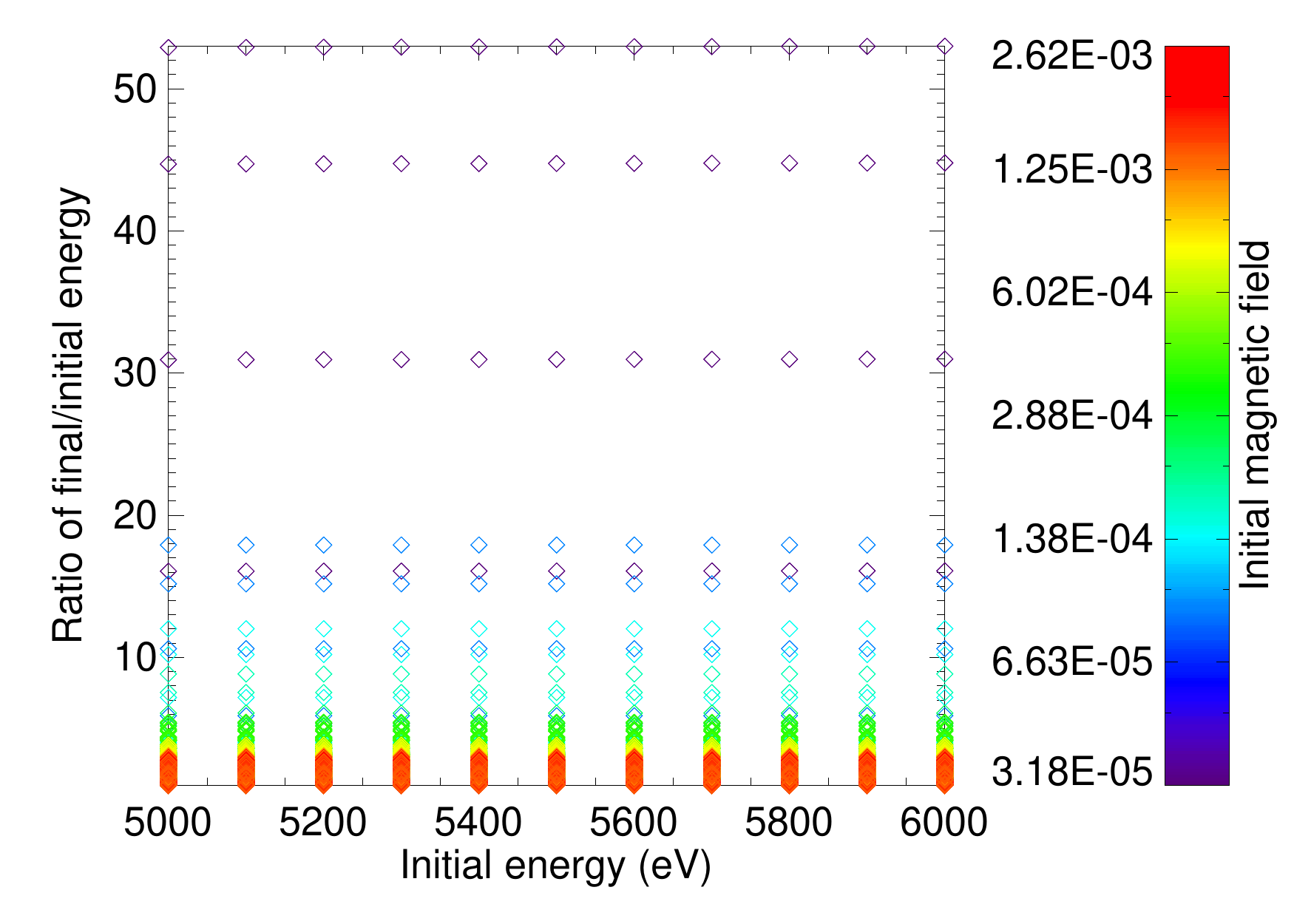}}
  \caption{Ratio of final to initial energy. Each point indicates a different test particle. Colours show, on logarithmic scale, the magnetic field strength at the initial positions of the orbits. One can see a clear trend that higher energy gains 
  are correlated with initial positions in weak field regions. The energy gain does, however, also depend on the initial pitch angle, with orbits with initial pitch angles close to $90^\circ$ 
  gaining more energy }
  \label{fig:ekin_ratio}
\end{figure}

Figure \ref{fig:ekin_ratio} shows the energy gain of particles as the ratio between the final and initial energy.
The values of initial energies chosen can be identified as the vertical bands on the graph. 
For the initial conditions investigated here, we find that the final energy can be up to $53$ times the initial energy (top boundary of Fig.\ \ref{fig:ekin_ratio}). 
Most particles ($98.5$ \% of the initial conditions shown) have modest energy gains of up to a factor of $10$.

Furthermore, for $2$ \% of the initial conditions shown, 
the particles lose energy compared to the initial state, but these are all particles which escape the trap almost immediately (within $1.5$ s in the normalisation discussed above). 
These particles all start outwith the central region of the CMT and usually have parallel velocities which take them directly to the nearest foot point of the field line they start on. 
Even some of the particles staying longer within the CMT are actually never really trapped, i.e. they do not mirror before crossing the lower boundary ($y=0$).  These particles usually have an initial parallel
velocity which takes them in the direction of the foot point further away from their starting position, which means they simply take longer to reach the lower boundary.
It is interesting that, despite not being trapped, even some of these particles gain energy because they encounter stronger magnetic field values while travelling to the point where they leave the CMT.

A closer investigation shows that the ratio between final and initial energy is determined very strongly by the initial position and the initial pitch angle, and only to a much lesser extent by the initial energy, at least
over the range of initial energies studied in the present paper ( we did carry out a limited number of test particle calculations for much smaller initial energies around a few $110$ eV, but found no qualitative
difference in the energisation process; higher initial energies would lead to final energies close to the electron rest mass energy and would thus require a relativistic calculation).
In particular, the initial position determines the initial magnetic field strength that the particle experiences. In Fig.\ \ref{fig:ekin_ratio} the magnetic field strength at the
initial position of the particles is indicated by the colour of the symbols, with the values being shown by the colour bar. 
As a general trend, particles starting in regions of 
lower magnetic field have the higher energy gains. 
Although the initial pitch angle is not indicated in Fig.\ \ref{fig:ekin_ratio}, we find that apart from starting in a region with lower magnetic field strength, the orbits
with the highest energy gains also have initial pitch angles which are closest to $90^\circ$, i.e. the particles have small initial parallel energies. The nearly horizontal bands seen in
Fig.\ \ref{fig:ekin_ratio} are actually made up of particle orbits which start at the same initial position with the same initial pitch angle, but different total initial energies.
We found that the particles with the largest energy gain ratio (above $30$) start in the centre of the trap (at $x=0$, $y=2.2$ L for those plotted here) with a pitch angle close to $90^\circ$.
These findings indicate that the betatron effect plays a major role for particles with the highest energy gains.

\subsection{The effect of the initial pitch angle}

\begin{figure}
   \resizebox{\hsize}{!}{\includegraphics{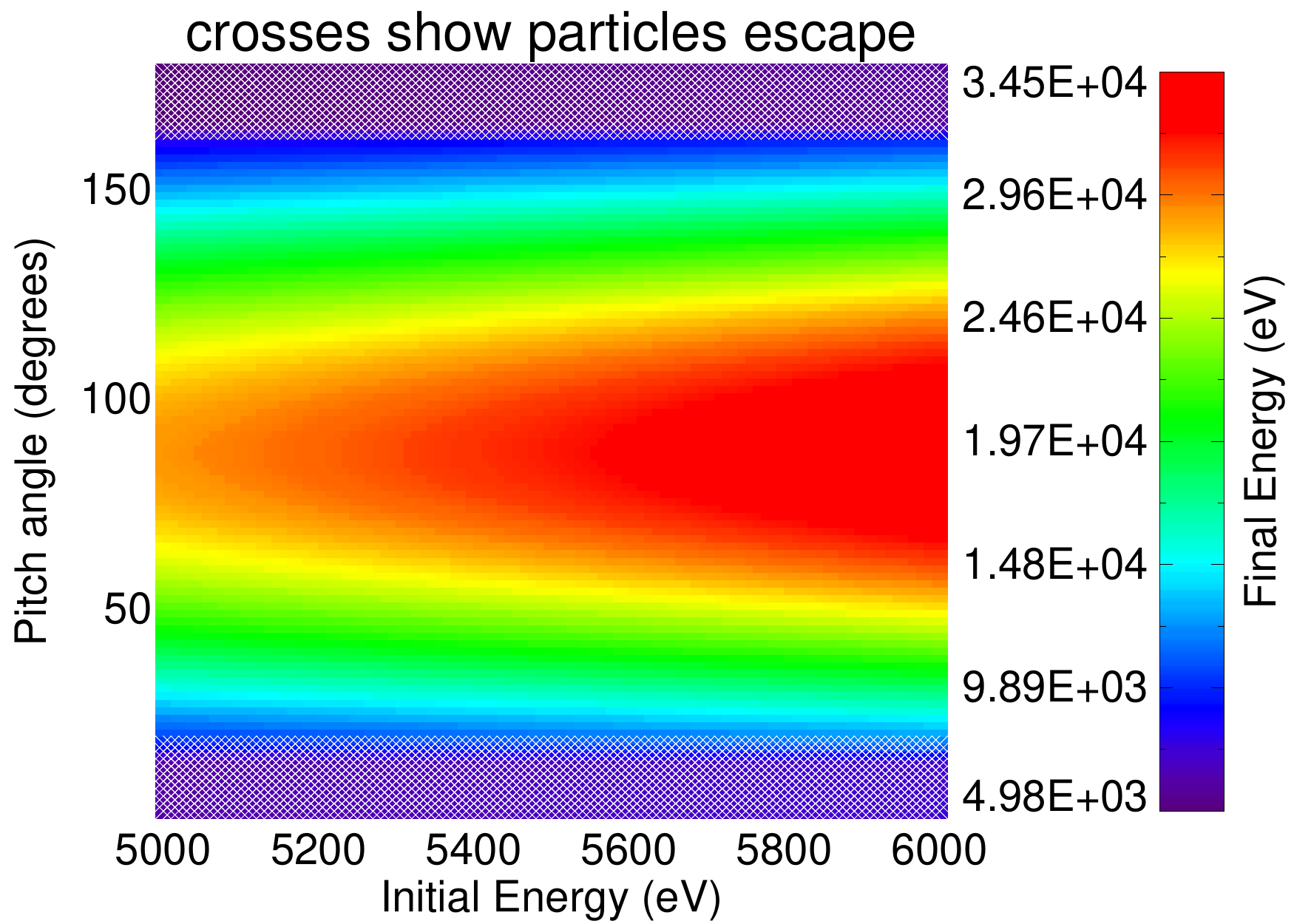}}

  \caption{Final energy (colour contours) of test particles with the same initial position ($x=0.1$ L, $y=2.0$ L), but different initial energies ($y$-axis) and pitch angles ($x$-axis). Crossed squares indicate 
  particles that escape before the trap has collapsed. For this initial position the highest energy particles have pitch angles closest to $90^\circ$}
  \label{fig:ekin_pitch_energy}
\end{figure}


To investigate more closely the effect of the initial pitch angle on the energy gain, we show in
Fig.\ \ref{fig:ekin_pitch_energy} the final energies (colour contours) for particle orbits starting at the same position ($x=0.1$ L, $y=2.0$ L), but with different initial energy ($x$-axis) and pitch angle ($y$-axis).
For these orbits the initial pitch angle varies between $1.8^\circ$ and $178.2^\circ$, 
and initial energy varies between 5keV and 6keV as shown. 
Particles that remain in the trap until the final time have
initial pitch angles between 19.6$^\circ$ and 162.2$^\circ$. 
Particles that  eventually escape the trap had initial pitch angles $\leq 17.8^\circ$ or $\geq 164.0^\circ$ in our grid of initial pitch angle values
 (both the pitch angle and the loss cone angle in our CMT model are time-dependent and therefore no general simple condition for particle escape can be given).
Escaping particles are indicated by crossed squares in Fig.\ \ref{fig:ekin_pitch_energy}.
The particles ending up with the highest final energy (about $34.5$ keV) have pitch angle closest to $90^\circ$ and start 
with the highest initial energy, consistent with the conclusions of the previous section.

\begin{figure*}
\centering
   \includegraphics[width=180mm]{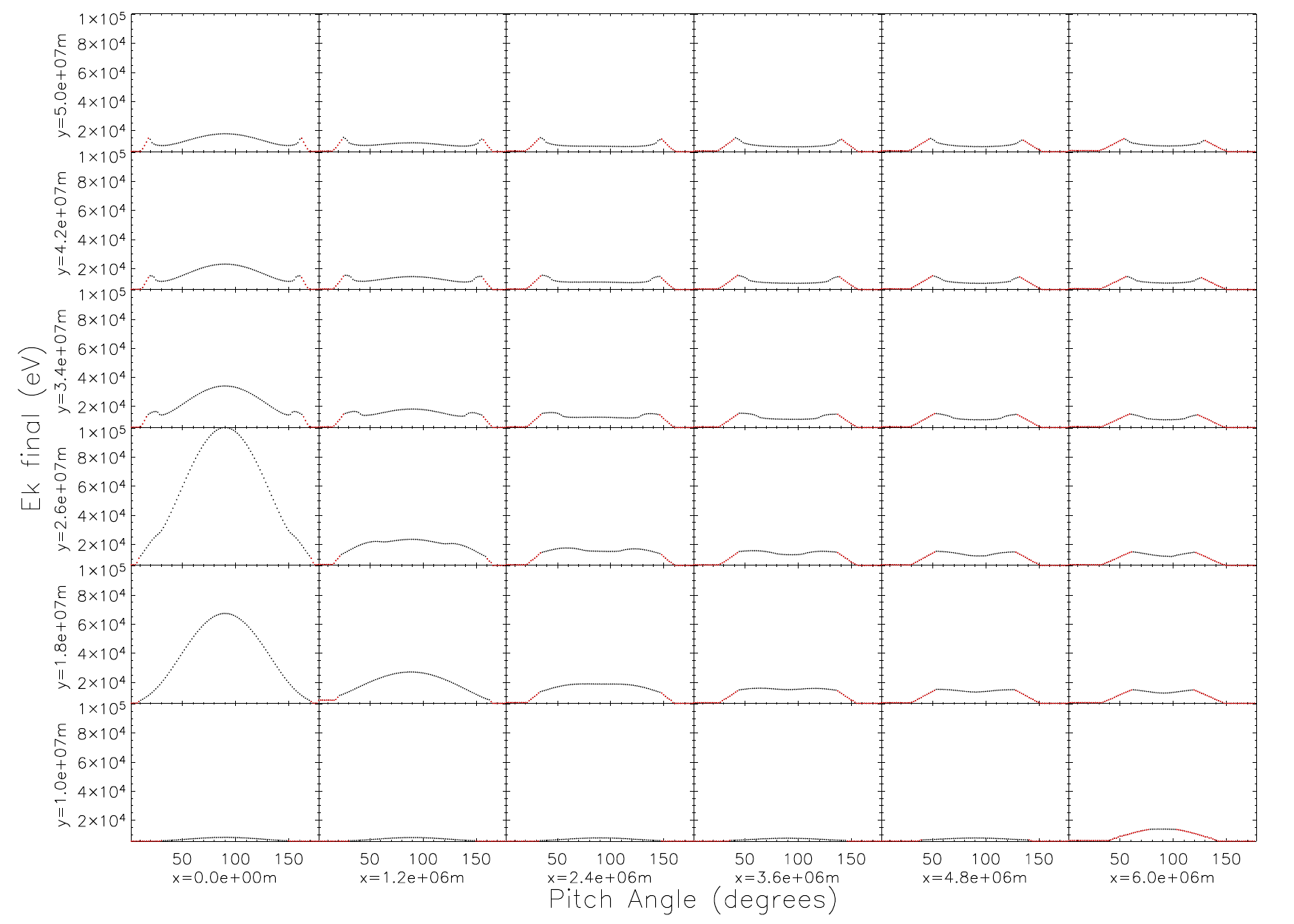}

  \caption{Final energy vs.\ pitch angle for different initial positions. All particles start with the same initial kinetic energy ($5.5$ keV). 
  Each plot is similar to a vertical cut through Fig.\ \ref{fig:ekin_pitch_energy}. 
  Red points indicate which particles escape. Note that the initial pitch angle leading to the maximum energy gain is not always $90^\circ$, but depends on initial position.}
  \label{fig:alpha_position}
\end{figure*}
The effect of varying the initial pitch angle at different initial positions, for a fixed initial energy of $5.5$ keV, is shown in Fig.\ \ref{fig:alpha_position}. The plots show final energy distributions versus initial pitch angle
for any combination of initial positions out of $x=0.0$, $0.12$, $0.24$, $0.36$, $0.48$ and $0.6$ L with $y=1.0$, $1.8$, $2.6$, $3.4$, $4.2$ and $5.0$ L.
Basically, every plot shown in Fig.\ \ref{fig:alpha_position} can be considered as a vertical cut through a figure similar to Fig.\ \ref{fig:ekin_pitch_energy} at $5.5$ keV for each of the initial positions.
In Fig.\ \ref{fig:alpha_position}, black dots indicate particle orbits which remain trapped, whereas red dots indicate escaping particle orbits.
It is obvious that particles with pitch angles deviating substantially  from $90^\circ$ escape from the trap more easily. It can also be easily seen that for initial positions further
away from the centre of the trap in the horizontal direction ($x$-direction), the range of initial pitch angles leading to escaping particle orbits becomes larger.

One can also see again that the particles with the highest final energy start with pitch angles close to $90^\circ$ in the centre of the CMT ($x=0.0$ L) and within the region of weak magnetic field (see e.g. $y=2.6$ L). 
While there is still a maximum of the final energy distribution around a pitch angle of $90^\circ$ in the CMT centre for other values of the initial height $y$, the value of the maximum energy is reduced compared 
to $y=2.6$ L. Another feature of the final energy as a function of pitch angle for increasing initial height $y$ is the development of secondary maxima at small and large pitch angles. A first indication is already visible
for $y=2.6$ L, but becomes increasingly clearer for larger initial $y$ values. The largest energy values of the secondary maxima occur close to the point of transition from trapped to escaping particle orbits. 
Similar trends as for the CMT centre at $x=0$ are also seen for the other values of $x$, although the final energies drop strongly in value. The maximum energy around the pitch angle of $90^\circ$ actually
turns into a local minimum, with the secondary maxima for small / large pitch angle becoming the highest energies as one moves away from $x=0$ at constant initial height $y$.
An explanation for these features is that the largest energy gains at the CMT centre are caused by the betatron effect, because the largest increase in magnetic field with time occurs at the centre of the CMT.
Particles starting close to the CMT centre with a pitch angle around $90^\circ$ stay very close to the CMT centre and thus basically gain all their energy through the betatron effect. Particles with small or large pitch angle have larger oscillation amplitudes for parallel and perpendicular energy whilst inside the trap. While the trap collapses, the particles move on field lines which shorten (by becoming less curved) and the distance between successive bounces becomes shorter. These
particles could therefore be mainly accelerated by the first order Fermi effect. This would explain the secondary peaks for smaller and larger pitch angles.

Particles starting away from the centre of the CMT do not experience the same large difference between initial and final magnetic field strength as the particles at the centre of the CMT, 
and thus the betatron effect becomes less efficient as the initial position moves away from the CMT centre. For small or large pitch angles, however, the Fermi effect could still operate, 
but only for particles outside the loss cone. This would explain the
final energy minimum and the secondary maxima for initial positions outside the centre of the CMT.

%

%
%
%
%

\section{Comparison of two particle orbits with different initial pitch angle} 
\label{sec:mechanisms}

\begin{figure}
\resizebox{\hsize}{!}{\includegraphics{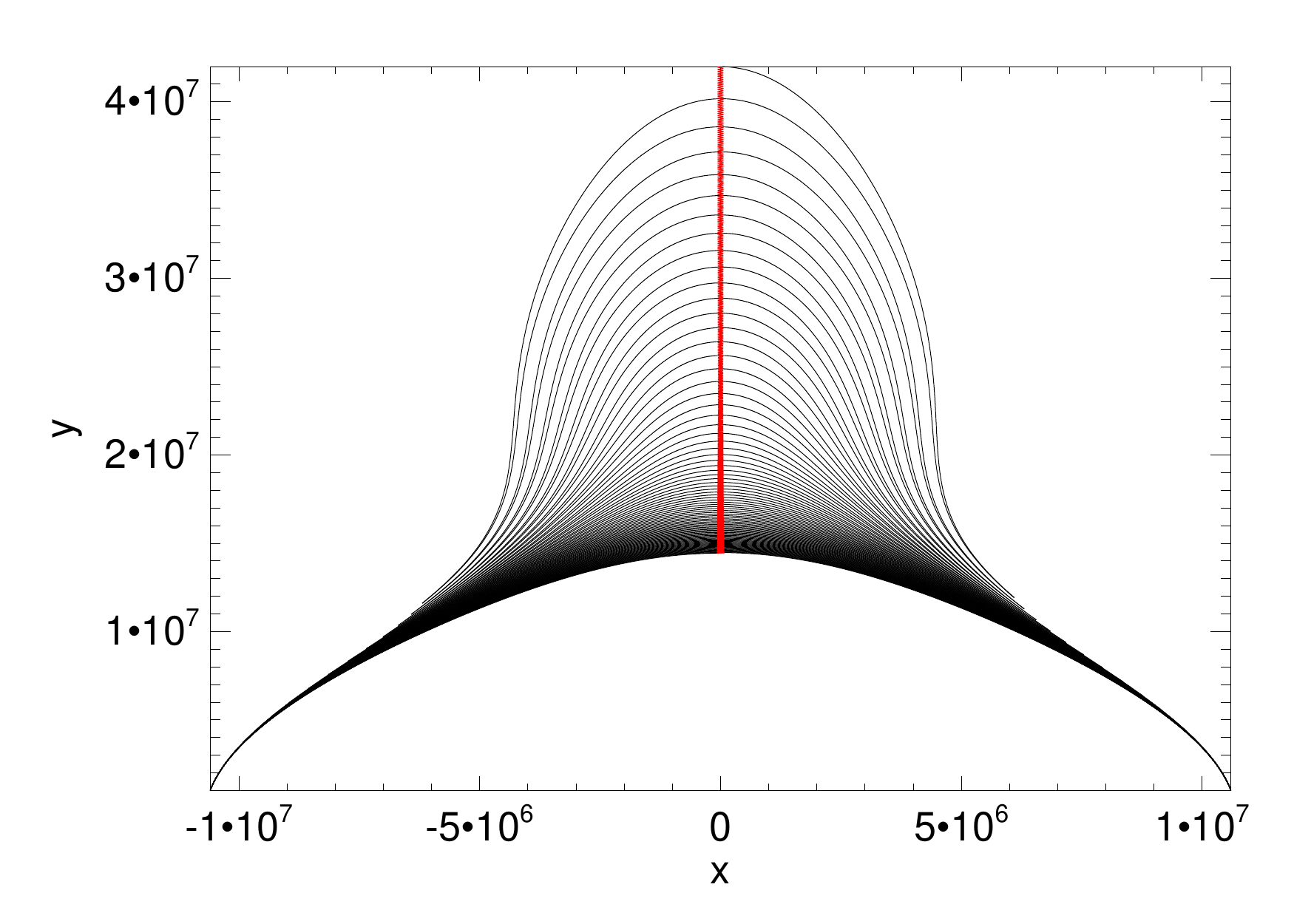}}
  \caption{The two test particle orbits. Particle orbit 1 with initial pitch angle $165.8^\circ$ is shown in black, particle orbit 2 with initial pitch angle close to $86.2^\circ$ is shown in red.}
  \label{fig:orbits1and2}
\end{figure}
To gain a better understanding of the different acceleration processes described above and how they depend on the initial pitch angle, we investigated in detail two particle orbits with the same initial position
 ($x=0$ L, $y=4.2$ L) and energy ($5.5$ keV), but with different initial pitch angles.
Particle orbit 1 has an initial pitch angle of $165.8^\circ$, i.e. the particle is moving initially mainly in the direction opposite to the field line. Particle orbit 2 has a pitch angle of $86.2^\circ$, 
so most of its initial energy is associated with the gyrational motion perpendicular rather than parallel to the field. Both particle orbits are shown in Fig.\ \ref{fig:orbits1and2}. 
As is to be expected, particle orbit 1 extends far along the field line, well into the legs of the trap, whereas particle orbit 2 remains close to the centre and mirrors more frequently. 
Both orbits remain on the same field line at all times.


\subsection{Time evolution of particle energies}

\begin{figure}
 \resizebox{\hsize}{!}{\includegraphics{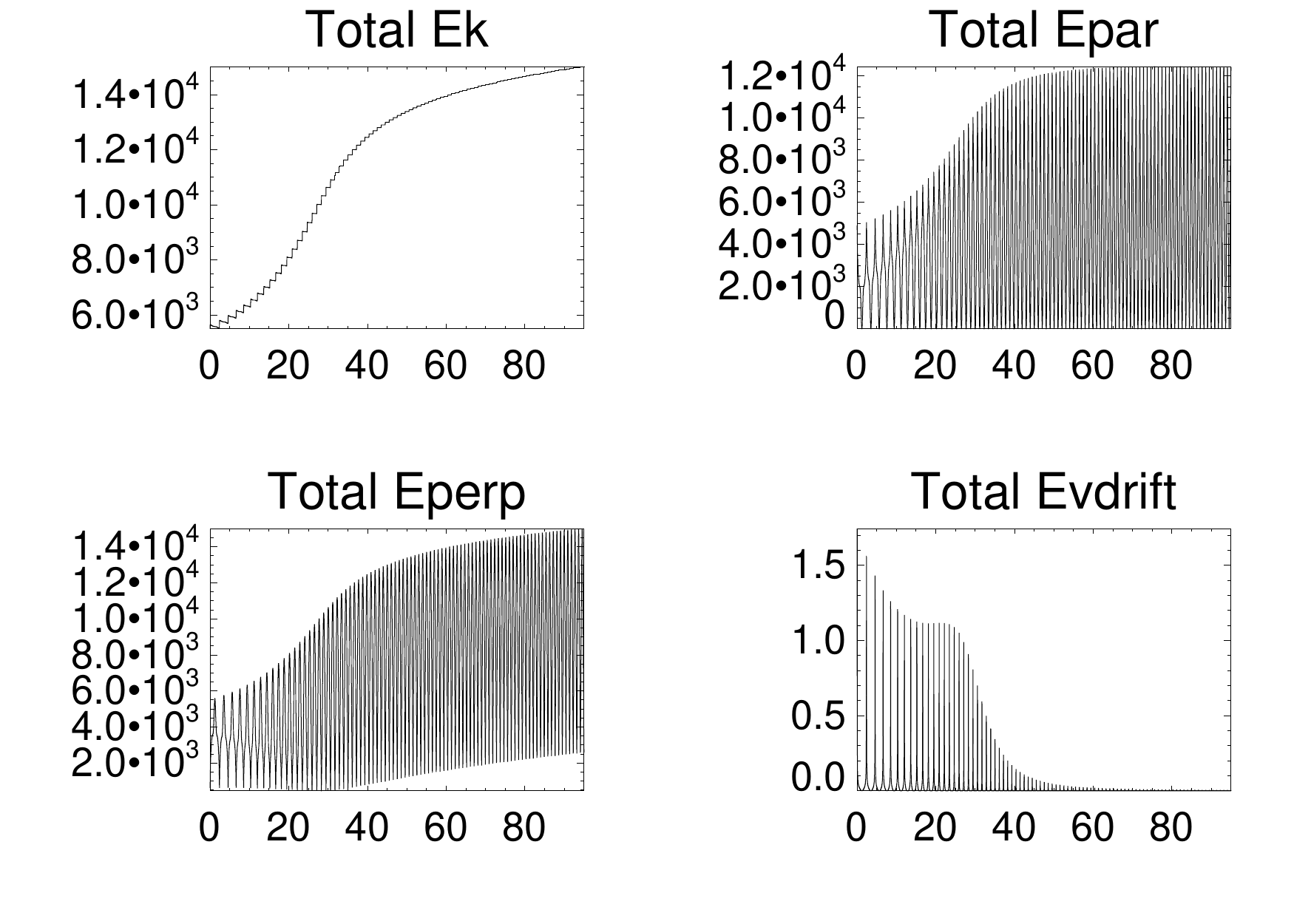}}
  \caption{Plots of time evolution of energy for particle orbit 1 (initial pitch angle $165.8^\circ$). Shown are the total kinetic energy (upper left panel), the parallel energy (upper right panel), 
  the perpendicular energy (lower left panel), and the energy associated with the $\mathbf{E}\times\mathbf{B}$-drift motion (lower right panel). In the normalisation discussed in the text, the numbers
  on the $x$-axis can be interpreted as seconds and the numbers on the $y$-axis as electron volts. }
  \label{fig:4energy_1}
\end{figure}
The upper left panel of Fig.\ \ref{fig:4energy_1} shows the time evolution of the total kinetic energy for orbit 1. 
The time evolution of the total energy for this orbit shows features which are very similar to the 
energy evolution of the particle orbit investigated by \citet{giuliani:etal05}. The energy increases in steps when the 
guiding centre moves along the top of the field line it is on, and it decreases slightly closer to the mirror points. As shown by \citet{giuliani:etal05}, the steps
are caused by the curvature  term in the parallel equation of motion and gives rise to an initial average increase in parallel energy ($mv^2_{\parallel}/2$). 
This is confirmed for particle orbit 1 by the plot of the parallel energy shown in the upper right panel of  Fig.\ \ref{fig:4energy_1}. 
A clear increase is visible when looking at the envelope of maxima of the parallel energy. These maxima occur when the particle passes through the centre of the trap ($x=0$), which
is consistent with the findings of \citet{giuliani:etal05}.
Obviously, for every trapped particle the minimum value of the parallel energy is zero (at the mirror points), but on average the parallel energy increases with time.

At the same time the perpendicular energy associated with gyrational motion of the particle ($MB$) also increases on average, as shown in the lower left panel of Fig.\ \ref{fig:4energy_1}.
The perpendicular energy has its maximum values at the mirror points and its minimum when passing through the centre of the CMT. However,
even at the centre of the CMT, the perpendicular energy is increasing with time. The increase of the perpendicular energy is clearly a consequence of the collapse of the magnetic field and the
corresponding increase in magnetic field strength along the particle orbit.

For comparison, we also show in the lower right panel of Fig.\ \ref{fig:4energy_1} the energy associated with the $\mathbf{E}\times\mathbf{B}$-drift motion, $m \mathbf{u}^2_E /2$, where
\begin{equation}
\mathbf{u}_E=\frac{1}{B^2} \mathbf{E} \times \mathbf{B}\mbox{.}
\end{equation}
Compared to the other parallel and perpendicular energies, the energy of the $\mathbf{E}\times\mathbf{B}$-drift motion is insignificant (here it is smaller by a factor of about $10^{-4}$).
Even at the initial time, the energy due to this drift is not significant when compared to the others, contributing $0.1\%$ at most. As the CMT collapses the energy associated with
$\mathbf{E}\times\mathbf{B}$-drift generally decreases to zero.

Obviously, due to the nature of trapped particle motion there is a constant interchange between parallel and perpendicular energy along any trapped particle orbit and the two
energy forms show the corresponding oscillations between maximum and minimum values. Naturally, these oscillations are out of phase and if added up lead to the total
energy not having any oscillation apart from the step-like behaviour discussed above. The average perpendicular energy and the average parallel energy are comparable 
for this orbit.

\begin{figure}
  \resizebox{\hsize}{!}{\includegraphics{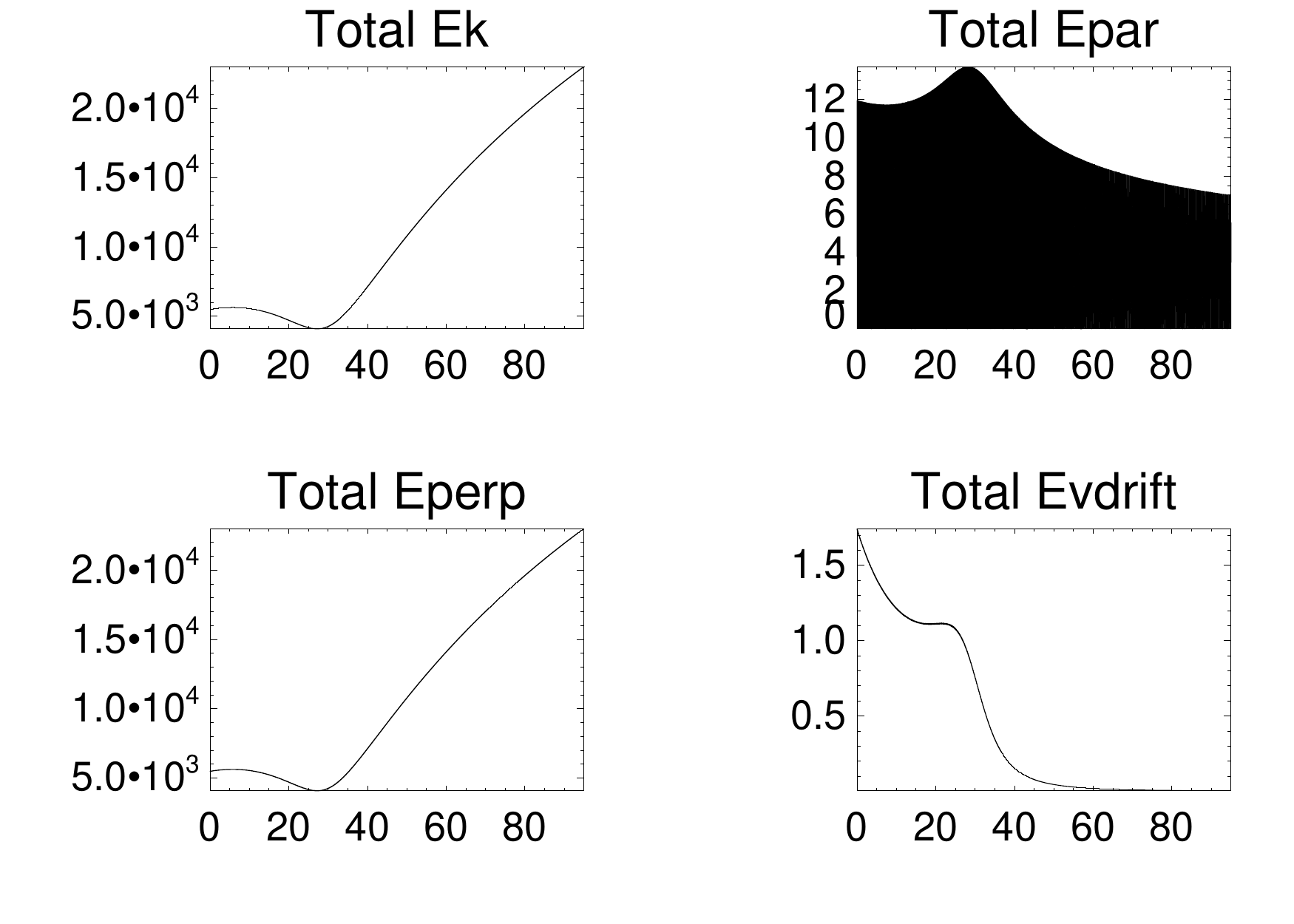}}
  \caption{Plots of time evolution of energy for particle orbit 2 (initial pitch angle $86.2^\circ$). Shown are the total kinetic energy (upper left panel), the parallel energy (upper right panel), 
  the perpendicular energy (lower left panel), and the energy associated with the $\mathbf{E}\times\mathbf{B}$-drift motion (lower right panel).
  Using the normalisation discussed in the text, the numbers
  on the $x$-axis can be interpreted as seconds and the numbers on the $y$-axis as electron volts.}
  \label{fig:4energy_2}
\end{figure}
For particle orbit 2, the energy shows a very different behaviour (see Fig.\ \ref{fig:4energy_2})
The total energy (upper left panel) again shows an overall gain, but only after some energy decrease at the beginning. The step-like behaviour, seen in the total energy for orbit 1, is not 
visible for orbit 2. 
The parallel energy (shown in the upper right panel) is again periodic, 
but this is more difficult to see as there are far more bounces 
due to the particle being trapped with mirror points very close to the centre of the CMT. 
It should also be noted that the parallel energy for this particle orbit is three orders of magnitude smaller than the total energy. 
This explains why we do not find the step-like behaviour seen for orbit 1, as it is simply too small to see on the scale of the total energy, although a closer investigation shows that
it is still present, but with a much smaller amplitude than for orbit 1.
We also remark that the peak seen in the parallel energy at about $28$ s corresponds to the minimum in total energy around the same time.

As the parallel energy is so much smaller than the total energy, it is clear that the perpendicular energy 
must be the dominating contribution to the total energy, and the two are indeed almost identical (see lower left panel). As for particle orbit 1, the energy associated with the 
$\mathbf{E}\times\mathbf{B}$-drift motion is negligible (see lower right panel).
However, a closer investigation shows, similar to the case of orbit 1, there are still small periodic variations in the perpendicular energy, although they are not visible on the scale shown here.
Because the bounce points are close to the centre of the CMT, the magnetic field does not change much over the period of a single particle oscillation, and thus
$E_\perp = M B$ does not change much either. An interesting feature of the perpendicular and the total energy time evolution is that there is an energy decrease to start with and that both energies increase
only after they have gone through a minimum. This feature can be explained quite easily by looking at the magnetic field structure of our CMT model. The CMT magnetic field strength has its minimum in the centre of the CMT at a height of about $y=2$ L at the beginning. Although the magnetic field evolves in time and the minimum in magnetic field strength eventually disappears, particles initially situated above this minimum and moving mainly downwards with collapsing field lines in the
centre of the CMT will pass through this minimum magnetic field strength region 
and their perpendicular energy will decrease accordingly. Once they have passed through that region the magnetic field 
will increase again and the perpendicular energy will increase as well, which is exactly what is seen in the two left panels of Fig.\ \ref{fig:4energy_2}. 
Generally, we can conclude that for particle orbits like orbit 2, the betatron effect is the dominating mechanism of energy gain.

As orbits 1 and 2 start at the same initial position, they must both pass through the field strength minimum, although orbit 1 will only pass through it when in the centre of the CMT, i.e. when its perpendicular energy is
at its minimum. A closer investigation does show that the graph of the perpendicular energy for orbit 2 has the same shape as the lower envelope of the perpendicular energy plot for orbit 1.
An indication of this can be found in the lower left panel of Fig.\ \ref{fig:4energy_1}.
More generally, any other particle orbit starting at the same position should have a perpendicular energy graph with a lower envelope of the same shape. This shape is determined by variation of the magnetic 
field strength $B$ with height at the centre of the CMT ($x=0$). The perpendicular energy graph for any particle is given by the product of $B$ and the magnetic moment, which is a constant in guiding centre
theory, and thus the minima of the perpendicular energy correspond to the minima in $B$ along an orbit.

\subsection{Longitudinal invariant and bounce length}

We already showed that the longitudinal invariant $J$ is very well conserved 
for the particle orbit investigated in \citet{giuliani:etal05}. This is also the case for the two orbits discussed above. For particle orbit 1 the maximum value is only
$0.037 \%$
larger than the minimum, and for particle 2 the maximum is only
$0.062 \%$
larger than its minimum.
\begin{figure*}
\centering
\includegraphics[width=8cm]{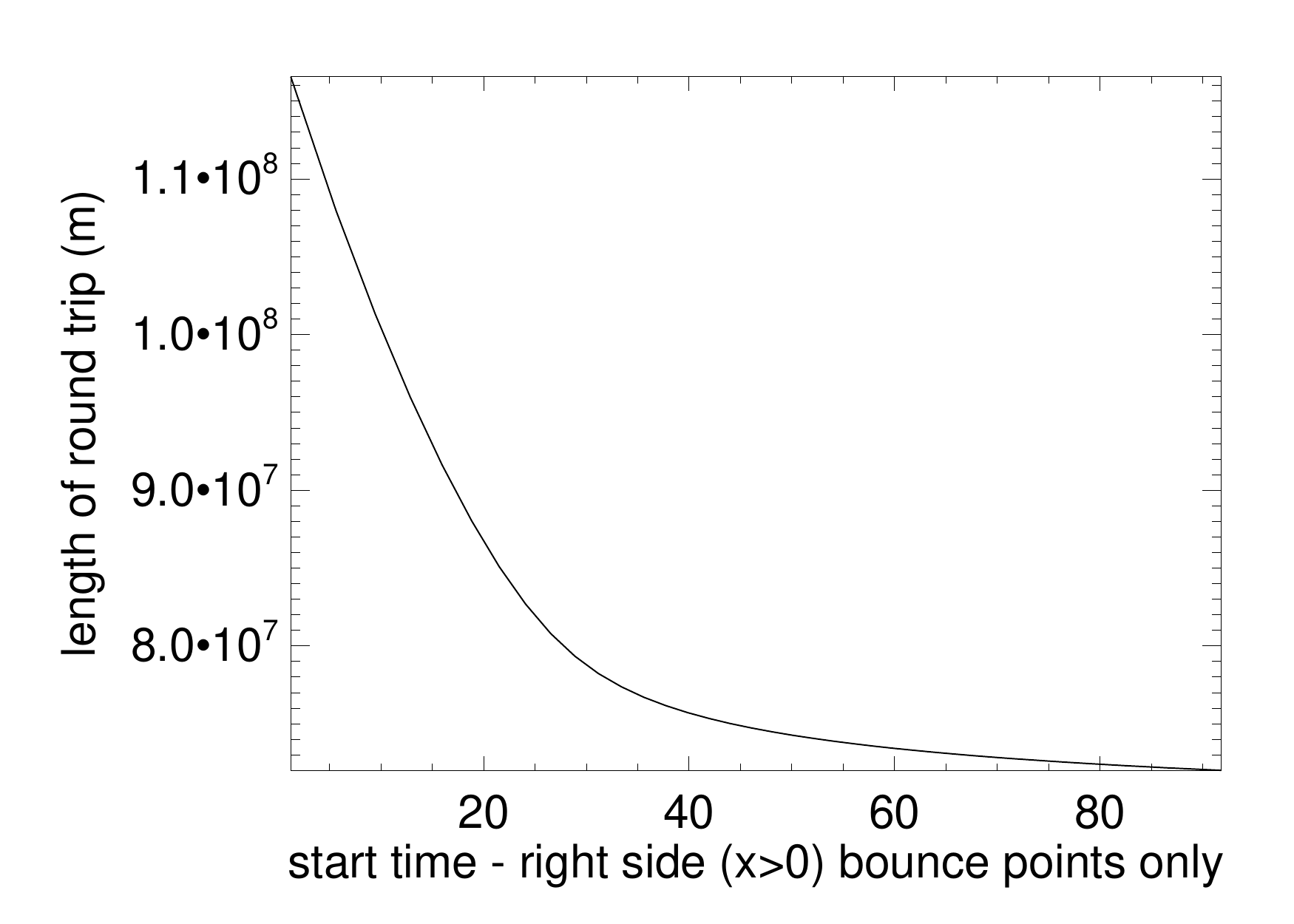}\includegraphics[width=8cm]{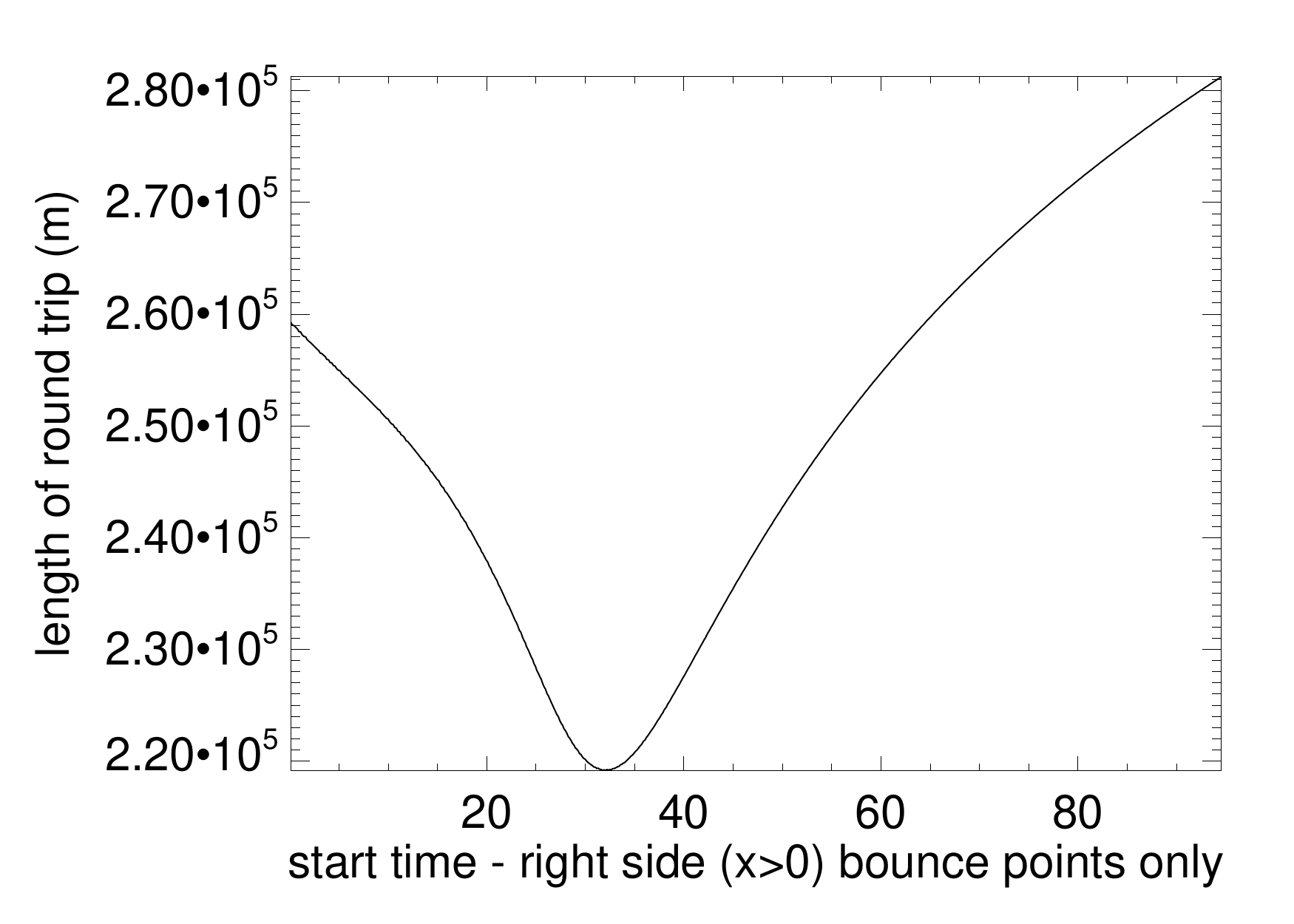}
  \caption{Bounce length as a function of time for particle orbit 1 (left panel) and 2 (right panel).}
  \label{fig:lengths}
\end{figure*}
Given that $J$ is a good invariant for the two orbits, an interesting question is how the distance between consecutive bounce points changes during the
evolution of the trap, because that could indicate the presence of the first order Fermi mechanism \citep[see e.g.][]{somov:kosugi97}.
The bounce lengths for the two orbits are shown in Fig.\ \ref{fig:lengths}. 
For particle orbit 1 the length decreases all the time while the trap is collapsing. This would be consistent with interpreting at least part of the energy gain 
as related to the first order Fermi mechanism. One should, however, bear in mind that, as discussed in detail by \citet{giuliani:etal05}, the parallel energy increases
mainly at the loop top due to the curvature term in the parallel equation of motion, as this gives rise to a source term in the parallel energy equation \citep[e.g.][]{northrop63}.
For particle orbit 2 the bounce length decreases to a minimum and then increases again. 
This is consistent with the increase of average  parallel energy at the beginning and decrease 
of the average parallel energy in the later stages of the collapse, as shown in the upper right panel of Fig.\ \ref{fig:4energy_2}.

\section{An asymmetric trap model}
\label{sec:asymmetric}

\subsection{Influence of initial conditions}
It is very unlikely that a solar flare would develop in perfect symmetry as we have assumed in the previous models. 
To investigate how the particle energisation processes change for an asymmetric CMT model, we make a simple modification to
the symmetric model used so far by placing the magnetic sources at different heights below the photosphere ($y=0$).
In particular, we now choose  $d_1=1$ and $d_2=1.5$ in Eq.\ (\ref{AGiuliani}). 
Using a larger value for $d_2$ means that the negative magnetic source on the right is farther below $y=0$ so the magnetic field above it at $y=0$ 
will be weaker than above the left source. 
Therefore,
We expect the particles to penetrate deeper into the the side with the weaker magnetic field.

\begin{figure}
\resizebox{\hsize}{!}{\includegraphics{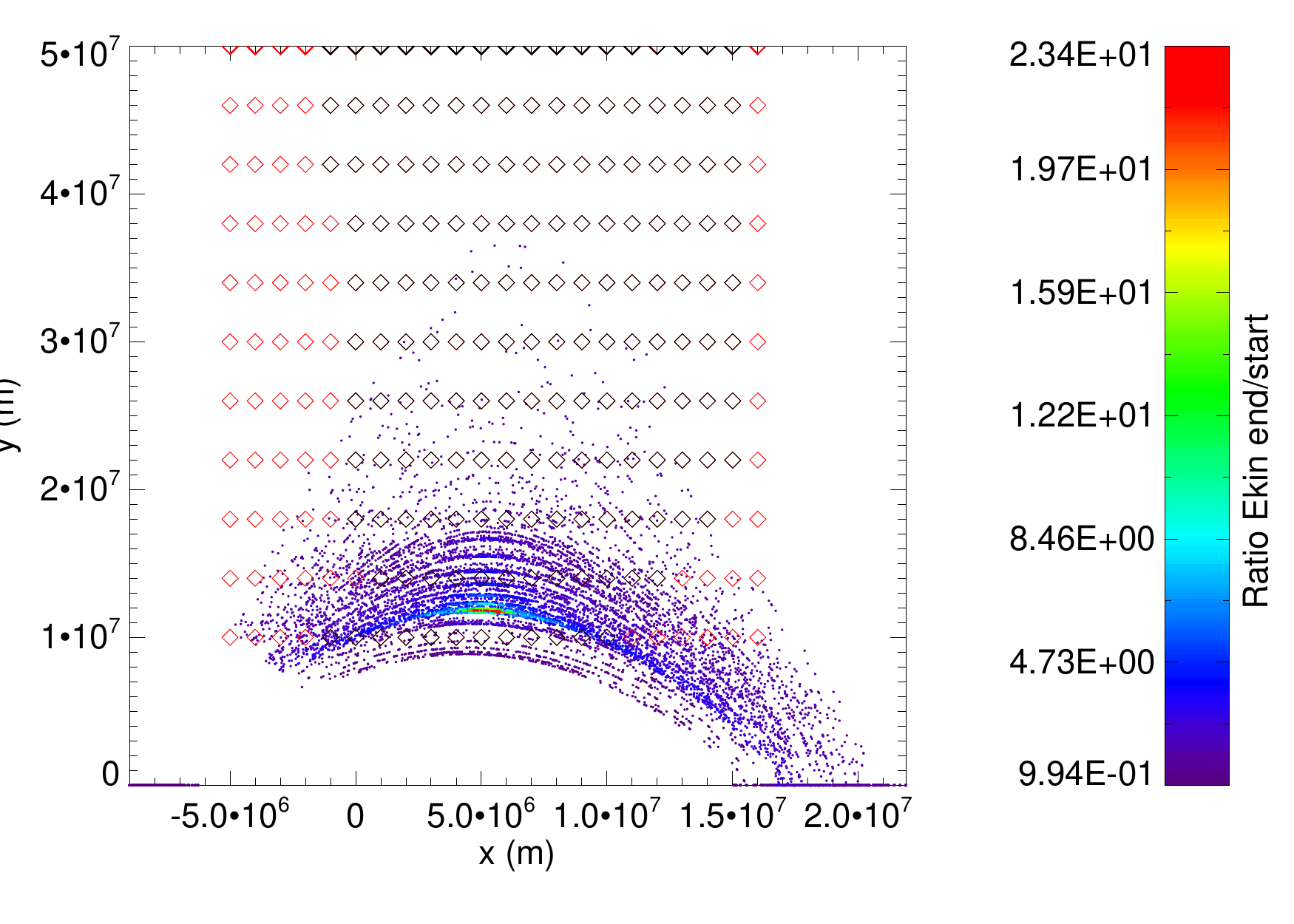}}
  \caption{Initial (diamond) and final (dot) positions of particles in the asymmetric trap. Black diamonds show that some particles starting there are trapped throughout. Colour of final position indicates energy gain.}
\label{fig:asym_initialpos}
\end{figure}

Because of the asymmetry of the CMT, it is no longer possible to clearly define the centre of the magnetic trap as a single unique value of $x$ as for the symmetric case with $x=0$.
We will therefore use the term ``central trap region'', which we regard as the region defined roughly by the mid-points between the two mirror points of trapped particles.
Due to the asymmetry, we have to use more initial positions in the $x$ direction to study the differences between particles starting on the left side and the right side of the trap. 
We keep the other initial condition ranges the same as the symmetric trap. We use a grid of  $22$ by $11$ equidistantly spaced initial positions for $-0.5 \mbox{ L} \le x \le 1.6 \mbox{ L}$ and 
$1 \mbox{ L} \le y \le 5\mbox{ L}$ (see diamond shaped symbols in Fig.\ \ref{fig:asym_initialpos}). As in the symmetric trap, for each initial position we calculate particle orbits for 
$11$ equally spaced values between $5$ keV and $6$ keV for the initial energy and 
$10$ values for the initial pitch angle between $16^\circ$ and $163^\circ$.
In Fig.\ \ref{fig:asym_initialpos} we also show the final positions of the particles remaining in the trap as dots, with the colour bar showing the energy gain. 
The highest energy trapped particles are still trapped in the region close to the loop top.

Again, many of the particles starting far away from the central trap region escape quickly. In Fig. \ref{fig:asym_initialpos} all the particles starting at a position
marked by red diamond escape from the trap before the final time.  
It may seem from  Fig.  \ref{fig:asym_initialpos} that the final positions of the particles seem to be generally higher than the final positions for the symmetric  trap model shown in Fig. \ref{initialpos}. However, 
this is only due to the fact that we have extended the region of starting positions in $x$ and hence some particles starting with larger initial $|x|$ values are located on field lines that extend to larger heights in the central trap region.

One can see that  as expected in this asymmetric trap model the trapped particles mirror closer to the right footpoint where the magnetic field is weaker. 
The escaping particles are indicated by the dots on the $x$-axis in Fig. \ref{fig:asym_initialpos}. We see that on the right hand side the locations where the particles
have crossed the $x$-axis is much wider than on the left-hand side, which is easily explained by the fact that the magnetic field is weaker at the right hand side foot points.


\begin{figure*}
\centering
\includegraphics[width=180mm]{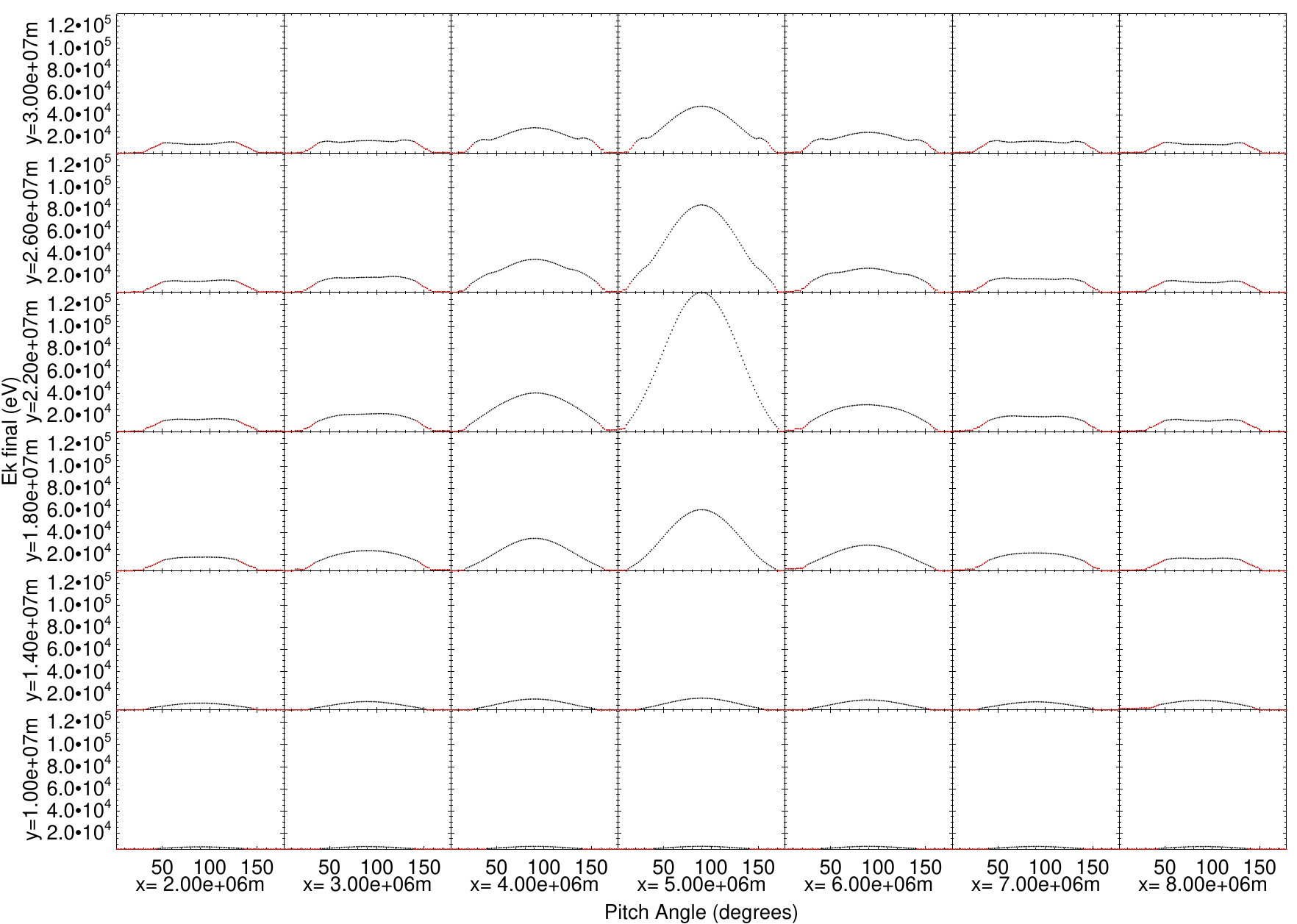}
  \caption{Final energy vs. pitch angle for different starting positions in the asymmetric trap. All particles start with energy 5.5keV. The region shown is centred on the maximum energy gain. The spacing is differnt to Fig.\ \ref{fig:alpha_position}. Red shows particles that escape before the trap has collapsed.}
\label{fig:asym_alpha_ek}
\end{figure*}
Another way to look at how the initial conditions can affect particle acceleration is to start particles with different pitch angles but all having the same initial energy from each position. This is shown in Fig.\ \ref{fig:asym_alpha_ek}. Each box shows particles starting at different positions in the trap. The energy is plotted at the final trap time or at the time of escape. Red points indicate escape from the trap. As with the symmetric trap, the maximum energy ratio is for particles starting near the trap centre with pitch angles close to 90$^\circ$. Also as in the symmetric trap, there is a secondary effect with `wings' on these graphs when the pitch angle is more field aligned and the starting position is further from the central trap region. From these points, the particles final enegy can be up to 131 keV, 24 times the initial energy.

\subsection{Energy gain in the asymmetric trap}
\begin{figure}
\resizebox{\hsize}{!}{\includegraphics{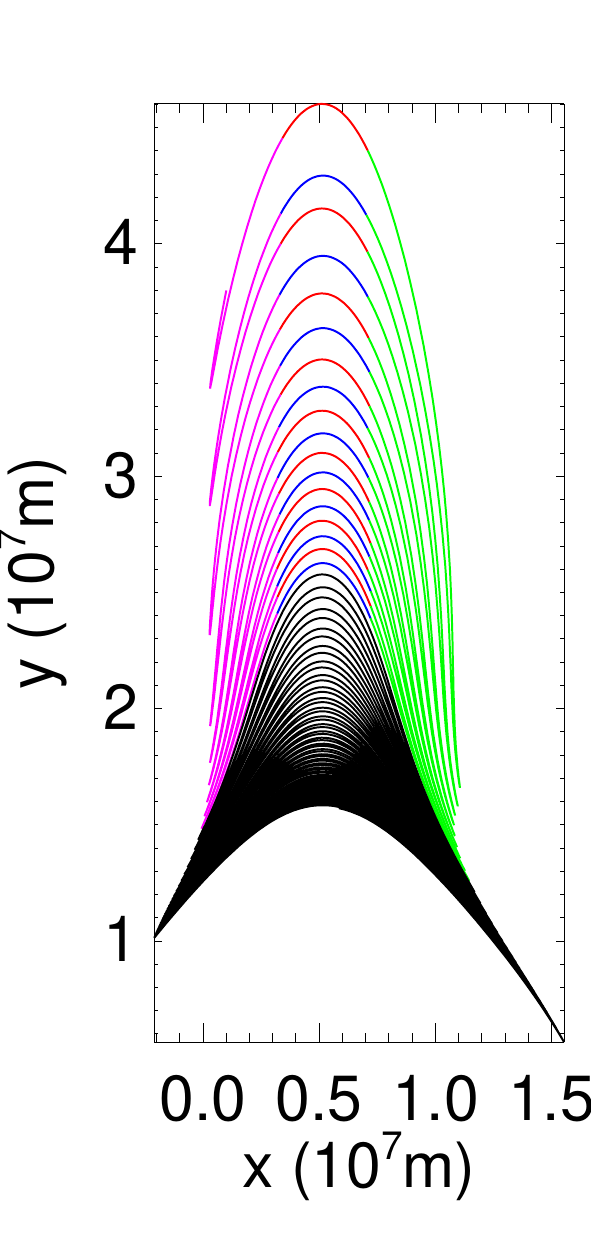}\includegraphics{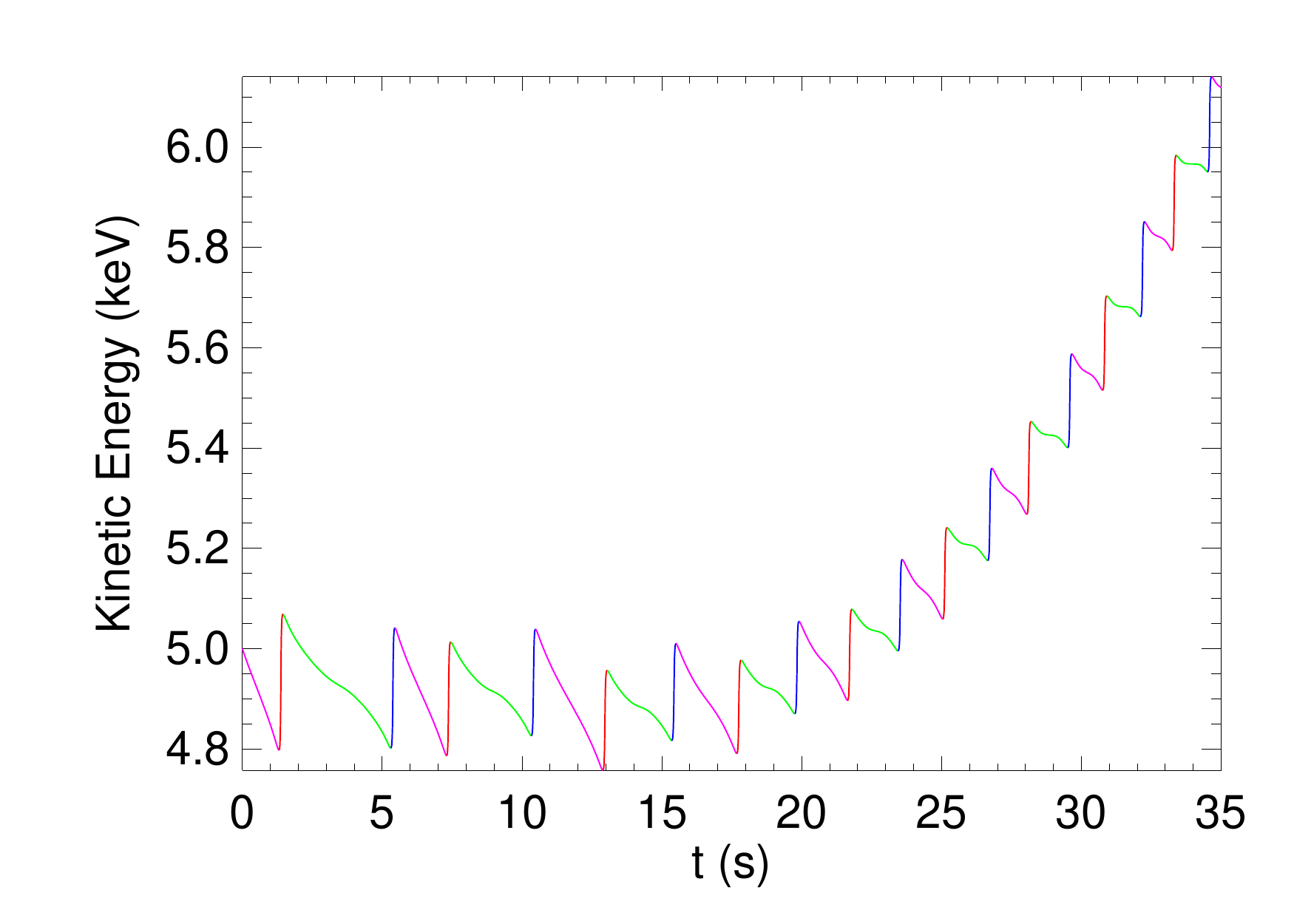}}
  \caption{Typical orbit and energy of a particle in the asymmetric trap. For the first 35 seconds, colours show corresponding parts where particles gain (red, blue) and lose (green, magenta) energy.}
\label{fig:asym_xyenergy}
\end{figure}
Asymmetries are also noticeable in the energy graph of individual particles in the trap.
Figure \ref{fig:asym_xyenergy} shows the complete orbit of a particle in the asymmetric trap in the left panel and the time evolution of its energy for the first 35 seconds in the right panel. 
We show only the first 35 seconds of the time evolution of the total energy, because after that time the step-like fashion of the energy increase and decrease phases is strongly reduced 
and it is more difficult to identify the different parts of the orbit. 
The part of the particle orbit after 35 seconds is indicated in black, for completeness.
This specific particle has initial energy $5$ keV and pitch angle $81.8^\circ$, starting position $x=0.1$, $y=3.8$, but the results for this particle are representative for most particles in the asymmetric case.

To illustrate the details of the particle energy increase and decrease along its orbit better, we present in Fig. \ref{fig:asym_xyenergy}  
the parts of the orbit in which the particle's energy increases and the parts where it decreases in a different colours. 
Specifically, the parts where the particle's energy increases are shown in red and blue, and the parts where its energy decreases are shown in green and magenta. 
The reason for using two different colours for the parts of the orbit where the particle
energy increases or decreases, respectively, is that this allows us to distinguish between 
the right- and left-hand side of the magnetic field configuration for the parts of the orbit where the energy decreases and between the different directions in which the particle
passes through the central trap region for the parts of the orbit in which the particle energy increases.

The boundaries of the different coloured sections are defined by the local minima and maxima of the total energy. 
Looking at the sections where the total energy decreases, we see that the particle energy decreases by a larger amount in the magenta sections (left hand side of the trap, i.e. stronger magnetic field)
than the green sections (right hand side of the trap, i.e. weaker magnetic field), although the particle remains in the parts of its orbit coloured in green for a longer time
than in the magneta sections. Here we ignore the initial magenta region in which the particle is located at the start. 

The left panel of Fig.\ \ref{fig:asym_xyenergy} showing the actual particle orbit provides some insight into these findings. 
We can see that the particle energy decreases along the parts of the orbit when the particle approaches the mirror points. As expected from the difference in magnetic field strength
in the asymmetric CMT model, the particle mirrors closer to the right hand side footpoint of the magnetic field line it is located on. This corresponds to the green part of the orbit (left panel) and 
of the energy graph (right panel) in Fig.  \ref{fig:asym_xyenergy}.  We see also that because the particle penetrates deeper into the magnetic field in the right hand side, it spends  longer in that part of the CMT than on the left hand side
(magenta section), where the mirror point is located higher up due to the higher magnetic field strength.
Also the particle energy decrease by a smaller amount than along the magenta section on the left. 

The particle gains energy only  when it is passing through the central trap region, shown in blue and red in the left panel of Fig.  \ref{fig:asym_xyenergy}.
This is consistent with the results for the symmetric CMT model, already discussed in some detail by \citet{giuliani:etal05}, who identified the terms related to the curvature
of the magnetic field lines in the equation for the parallel velocity as being responsible for the energy increase of the particles.
At first sight, the fact that the particle energy increases in the central  region of the trap is difficult to reconcile with a simple interpretation  of the energy gain in 
terms of Fermi acceleration model, which is usually associated with energy gain when the particle bounces off a moving obstacle \citep[see e.g.][]{longair2:94}, i.e. in a CMT
close to the mirror points.
However, the Fermi mechanism discussed for CMTs is connected to conservation of the second adiabatic invariant $J$, which is an integral over one complete
bounce period of a particle orbit. Therefore any energy gain process related to this adiabatic invariant can only be understood as operating in an average sense
over a complete bounce period of the particle.

Generally we conclude that while there are some quantitative differences between the energy gain and loss process in the symmetric and asymmetric CMT models, we 
did not find any significant qualitative differences.

\section{Summary and Discussion}
\label{sec:summary}

We have presented a detailed study of the particle energization processes in CMTs, using in particular  the symmetric CMT model of \citet{giuliani:etal05} and a modified asymmetric CMT model based on
the same theoretical framework.
We found that in the particular symmetric CMT model we studied, particle energies can increase by factors of up to approximately $50$, but that most particles experience a more modest energy increase.
While the energy increase does not depend strongly on initial energy, it does depend on the initial position of particles in the CMT and on the initial pitch angle.
Particles with the highest increase in energy start in the region of the CMT which initially has the smallest magnetic field strength and usually have pitch angles close to $90^\circ$.
The energy increase for these particles is caused mainly by the betatron effect as the trap collapses and the magnetic field strength along the orbit increases. Due to their
pitch angle these particles remain trapped close to the centre of the trap, which means that at the end of the CMT collapse the highest energy population of particles 
is confined in a region at the top of the most collapsed magnetic loop. This is consistent with previous results using other CMT models \citep[e.g.][]{karlicky:kosugi04}.

We also found that for particles with initial pitch angles differing substantially from $90^\circ$, but outside the loss cone at any time during the collapse of the CMT, 
a substantial increase in parallel energy is possible. On a superficial level this could be interpreted as first order Fermi acceleration as usually the distance
between mirror point is decreasing during the CMT evolution. A more careful investigation, however, corroborates the finding of \citet{giuliani:etal05} that
the particle energy increase is due to the curvature of field lines in the centre of the CMT. 
 In particular, we found the increase in parallel energy is caused by the curvature term in the parallel equation of motion, which in our model takes on its maximum value
in the center of the CMT and not close to the mirror points. 
The Fermi mechanism discussed for CMTs is 
associated with the conservation of the second adiabatic invariant $J$, which is an integral over one complete
bounce period of a particle orbit and thus should be understood as operating in an average sense
over a complete bounce period of the particle.

We found similar results for the asymmetric CMT model. While there are noticeable quantitative differences in the general appearance of
the particle orbits that can be easily explained by the magnetic field asymmetry, as well as some differences in the maximum total energy increase
that is possible, the qualitative features of particle energisation in the asymmetric CMT model were found to be the same as for the symmetric CMT model.



In view of recent findings that high energy radiation from loop tops or above loop tops \citep{masuda:etal94} is 
more common during solar flares than previously thought \citep[see e.g.][for an excellent review]{krucker:etal08},
it is tempting to associate the fact that the highest energy particles are trapped at the top of the loop with hard X-ray loop top sources. 
This has been suggested in the past by other authors
\citep[e.g.][]{somov:kosugi97, karlicky:kosugi04,karlicky:barta06,minoshima:etal10} and attempts have been made at 
calculating the characteristics of the hard X-ray emission 
expected from CMT models \citep[e.g.][]{karlicky:barta06}. 
In order to assess this properly, the present CMT models should be amended to include collisions with a 
background plasma along the lines of previous models for hard X-ray loop top emission using
static loop models \citep[e.g.][]{fletcher95,fletcher:martens98}. 
Apart from being a possible explanation for sources of coronal high energy radiation, the trapping of high energy 
particles in the corona can also contribute to the explanation of observations of
microwave emission from flaring loops \citep[e.g.][]{melnikov:etal02}.

Obviously, like many other models, the CMT model used in the present paper is highly simplified. Apart from the introduction 
of collisions with a thermal background plasma, a number of other improvements
should be made in the future. We have shown in this paper that  the curvature term in the parallel equation of motion plays an important role in 
the particle acceleration process in CMTs. For any 2D CMT model, the 
curvature drift (as well as, for example
the gradient-B-drift) is actually directed into the invariant direction (i.e. along the $z$-direction in our coordinate system). 
It turns out that the particle orbits do not move too far in the $z$-direction compared
to their motion in the $x$-$y$-plane, 
but it nevertheless raises the 
question whether the results would change for  a 3D CMT model. \citet{grady:neukirch09b} have recently presented a generalised theory allowing for 
3D CMT models including both 3D
magnetic fields and 3D flows, which could be used in the future.

As in the present paper the flow field associated with CMT models is usually assumed to be laminar. It is, however, 
highly unlikely that a violent event such as a solar flare will give rise to such 
regular behaviour. A possible improvement for future CMT models might be to add turbulent motion (and the corresponding electromagnetic fields)
onto the overall laminar motion associated with the collapse. A possible way of dealing with
this is to add a stochastic scattering term to the equations of motion, similar in principle, but different in detail, to a 
Coulomb collision term. There are several interesting questions that arise
in connection with such an approach, for example: How would the particle energization in a turbulent CMT 
change compared to a laminar CMT? How would the energy density associated with the turbulent flow
and EM fields evolve in a CMT? An interesting aspect of such models would be that they could provide a 
link between stochastic particle acceleration models \citep[e.g.][]{miller:etal97} and the standard
flare scenario in a similar way as proposed by e.g. \citet{hamilton:petrosian92}, \citet{park:petrosian95}, \citet{petrosian:liu04} and more recently by \citet{liu:etal08}.

\begin{acknowledgements}
We would like to thank the referee for helpful and constructive comments.
The financial support by the UK's Science and Technology Facilities Council is gratefully acknowledged.
\end{acknowledgements}

\bibliographystyle{aa}

\bibliography{aa-paper-revised}

\end{document}